# Generalized Delta Functions and Their Use in Quantum Optics

R.A. Brewster and J.D. Franson
University of Maryland Baltimore County, Baltimore, MD 21250 USA

**Abstract.** The Dirac delta function $\delta(x)$ is widely used in many areas of physics and mathematics. Here we consider the generalization of a Dirac delta function to allow the use of complex arguments. We show that the properties of a generalized delta function are very different from those of a Dirac delta function and that they behave more like a pole in the complex plane. We use the generalized delta function to derive the Glauber-Sudarshan P-function, $P(\alpha)$, for a Schrödinger cat state in a surprisingly simple form. Aside from their potential applications in classical electromagnetism and quantum optics, these results provide insight into the ability of the diagonal P-function to describe density operators with off-diagonal elements.

## 1. Introduction

The Dirac delta function $\delta(x)$ is an invaluable tool in many areas of physics, including electromagnetism, quantum optics, and field theory. Despite its commonly-used name, the Dirac delta function is only defined inside an integral and it is thus a distribution or generalized function [1-3] rather than a true function. Roughly speaking, $\delta(x)$ is singular at a point on the real axis and the integral of a function multiplied by the Dirac delta function will give the value of the function at that point as illustrated in figure 1a. Here we consider a generalization, $\tilde{\delta}(z)$, of the Dirac delta function (also a distribution) whose argument is allowed to be a complex variable. As a result, its properties are different from a conventional Dirac delta function and an integral along the real axis can give the value of the function at a point in the complex plane, which is similar in some respects to a contour integral around a pole as illustrated in figure 1b.

Generalized delta functions provide a convenient way to describe the singular nature of certain quasiprobability distributions, which are widely used in quantum optics and other areas of quantum physics [4-11]. Unlike a true probability distribution, quasiprobability distributions can have negative values that demonstrate the nonclassical properties of the corresponding system. As a result, plots of quasiprobability distributions can provide insight into the properties of a system in a mixed state. In addition, quasiprobability distributions can be used to calculate the properties of quantum systems that would be difficult to calculate in other ways.

The Glauber-Sudarshan P-function $P(\alpha)$ is a quasiprobability distribution that is especially useful for calculating the density matrix of a system using the relation

$$\hat{\rho} = \int_{-\infty}^{\infty}\int_{-\infty}^{\infty} P(\alpha) |\alpha\rangle\langle\alpha| \, d\alpha_r d\alpha_i. \tag{1}$$

Here $|\alpha\rangle$ is a coherent state with complex amplitude $\alpha$ with real and imaginary parts $\alpha_r$ and $\alpha_i$. The P-function, implicitly defined by equation (1), corresponds to a diagonal representation of the density operator in a basis of coherent states [4-6], whereas the density operators for nonclassical states have off-diagonal elements that are frequently as large as their diagonal elements. The P-function can be highly singular as a result, in which case it can only be represented as a distribution [1-3,12] rather than a true function [5,13]. Here we use the properties of the generalized delta function to calculate $P(\alpha)$ for a Schrodinger cat state in a remarkably simple form. This provides a straightforward explanation of how the diagonal P-function can represent off-diagonal density operators.

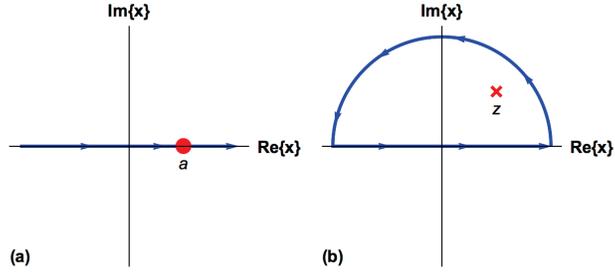

**Figure 1.** Comparison of a generalized delta function with a conventional Dirac delta function. (a) Integration of a function $f(x)$ multiplied by a conventional Dirac delta function $\delta(x-a)$ along the real axis gives the value of the function at the point $x=a$. The only singularity in $\delta(x-a)$ is at that point. (b) Integration of a function multiplied by a generalized delta function $\tilde{\delta}(x-z)$ along the real axis gives the value of the function at the point $x=z$ which may lie in the complex plane. This is similar to the properties of a contour integral around a pole where the path can be extended to infinity to form a closed loop. In addition, $\tilde{\delta}(x-z)$ is nonzero and singular over an extended region along the real axis.

   The main goals of this paper are to generalize the Dirac delta function to include complex arguments and then use it to calculate highly singular quasiprobability distributions. Generalized delta functions with complex arguments are defined and their properties derived in the next section using the theory of distributions, or generalized functions [1-3]. Section 3 briefly reviews quasiprobability distributions and some of their applications, including the relationship between the most commonly used quasiprobability distributions. The Glauber-Sudarshan P-function for a Schrödinger cat state is derived in section 4, which illustrates the usefulness of the generalized delta function as well as providing a simple form for $P(\alpha)$. As a further example, the effects of a linear amplifier on a Schrödinger cat state are considered in section 5, where it is shown that the generalized delta function arises naturally in the limit of low gain and decoherence. A summary and conclusions are provided in section 6.

## 2. Generalized Delta Functions

The main goal of this section is to define a generalized delta function $\tilde{\delta}(z)$ in such a way that

$$\int_{-\infty}^{\infty} f(x)\tilde{\delta}(x-z_0)dx = f(z_0), \qquad (2)$$

where $z_0$ is a point in the complex plane. We will show that the allowed test functions $f(z)$ must be an analytic function of the complex variable $z$ throughout the complex plane, which is not a requirement for an ordinary Dirac delta function. The test functions $f(z)$ for which Eq. (2) holds are also assumed to decrease sufficiently rapidly at large distances along the real axis that $f(x+iy)$ is $O(|x|^{-N})$ as $|x| \mapsto \infty$ for all values of $N$. This condition can be relaxed in certain situations, but it will be shown to be satisfied by all of the functions that are typically of interest in quantum optics.

   The sifting property of Eq. (2) is the same as that of an ordinary Dirac delta function aside from the fact that the argument can be complex. There are some important differences, however. As illustrated in Fig. 1, the integral of $f(x)$ multiplied by the generalized delta function will give the value of the function at the point $z_0$ in the complex plane, which is similar in some respects to a contour integral around a pole. The generalized delta function $\tilde{\delta}(x-z)$ is also nonzero and singular over an extended range of the real axis, unlike a Dirac delta function, as is discussed in more detail in appendix A.

Delta-functions with a complex argument have been briefly discussed previously [14-16] but with several different definitions and no rigorous proof of their properties. For example, delta functions with complex arguments have also been defined [16] as

$$\delta(x+iy) \equiv \delta(x)\delta(y), \tag{3}$$

where $x$ and $y$ are the real and imaginary parts of $z$. As discussed in more detail in appendix B, this definition of $\delta(z)$ corresponds to the product of two conventional Dirac delta functions, which is completely different from the properties of the generalized delta function in equation (2). One of the goals of this paper is to eliminate this ambiguity along with deriving the properties of $\tilde{\delta}(z)$. To avoid confusion, we will use the notation $\delta^2(x+iy) \equiv \delta(x)\delta(y)$ [4] to distinguish this from $\tilde{\delta}(z)$.

*2.1 Definition and basic properties*

The sifting property of Eq. (2) can be derived using the theory of distributions, or generalized functions [1-3]. Consider a sequence of functions $\{\phi_j(z)\}$ given by

$$\phi_j(z) = \frac{1}{\sqrt{2\pi}\sigma_j} e^{-z^2/2\sigma_j^2}, \tag{4}$$

where $\sigma_j = 1/j$. Equation (4) is equivalent to the one of the sequences of functions that are commonly used to define the conventional Dirac delta function, except that here $z$ can be a complex number. We will also consider the function $F_j(z_0)$ defined by

$$F_j(z_0) \equiv \int_{-\infty}^{\infty} f(x)\phi_j(x-z_0)dx. \tag{5}$$

Our goal is to show that $F_j(z_0) = f(z_0)$ in the limit $j \to \infty$.

The derivation is based on the contour integral shown in Figure 2. Since $f(z)$ and $\phi_j(z)$ are both analytic throughout the complex plane, the contour integral around any closed path is zero and

$$I_1 + I_2 + I_3 + I_4 = 0. \tag{6}$$

Here $I_1, I_2, I_3,$ and $I_4$ denote the integral of $f(z)\phi_j(z-z_0)$ along the four paths shown in Figure 2. $I_2$ and $I_4$ can be readily shown to vanish in the limit $L \to \infty$, where $L$ is the half-length of the integral along the real axis. As a result

$$F_j(z) = I_1 = -I_3 = -\int_{\infty}^{-\infty} f(x+ib)\phi_j[(x+ib)-(a+ib)]dx, \tag{7}$$

where $a$ and $b$ denote the real and imaginary parts of the point $z_0$. Equation (7) can be rewritten as

$$F_j(z) = \int_{-\infty}^{\infty} f(x+ib)\phi_j(x-a)dx. \tag{8}$$

It can be seen that the argument of $\phi_j$ in Equation (8) is now real.

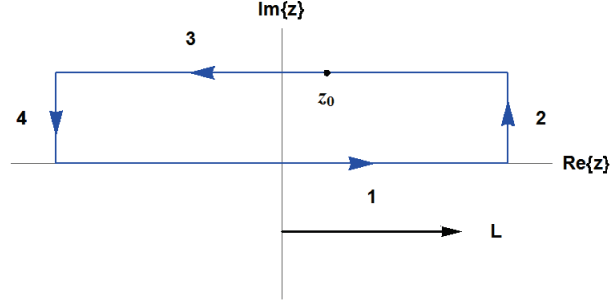

**Figure 2**. Derivation of the sifting property of a generalized Dirac delta function in Eq. (2) using integration around a closed contour that encloses the point $z_0$. Since the integrand is analytic in this region, the sum of the four integrals $I_1, I_2, I_3,$ and $I_4$ along the paths indicated in the figure is zero.

In order to prove the desired result, consider the quantity $q_j$ defined by

$$q_j \equiv |F_j(z_0) - f(z_0)|. \tag{9}$$

From Equation (8), this can be rewritten as

$$\begin{aligned} q_j &= \left| \int_{-\infty}^{\infty} f(x+ib)\phi_j(x-a)dx - f(z_0) \right| \\ &= \left| \int_{-\infty}^{\infty} [f(x+ib) - f(a+ib)]\phi_j(x-a)dx \right|. \end{aligned} \tag{10}$$

The last expression in Equation (10) follows from the fact that the integral of $\phi_j$ along the real axis is normalized to unity and $f(z_0)$ is a constant. Using Equation (10) and the Taylor series expansion of $f(z)$ gives an upper bound [3] on $q_j$:

$$q_j \leq Max|f'| \int_{-\infty}^{\infty} |x-a|\phi_j(x-a)dx. \tag{11}$$

The integral in Eq. (11) vanishes in the limit $j \to \infty$, which shows that $q_j = 0$ and $F_j(z_0) = f(z_0)$ in the same limit.

We can now define the generalized delta function $\tilde{\delta}(z)$ by

$$\int_{-\infty}^{\infty} f(x)\tilde{\delta}(x-z_0)dx \equiv \lim_{j \to \infty} \int_{-\infty}^{\infty} f(x)\phi_j(x-z_0)dx. \tag{12}$$

The fact that $F_j(z_0) = f(z_0)$ as $j \to \infty$ proves the desired sifting property of the generalized delta function in Eq. (2).

The generalized delta function is only defined within an integral and for analytic functions $f(z)$ with the properties described above. Thus $\tilde{\delta}(z)$ is a distribution, or generalized function, and not a true function as is also the case for the conventional Dirac delta function. The symbolic notation

$$\tilde{\delta}(z) = \lim_{j \to \infty} \phi_j(z) = \lim_{\sigma \to 0} \frac{1}{\sqrt{2\pi}\sigma} e^{-z^2/2\sigma^2} \tag{13}$$

is often used in the literature, even though the limit in Eq. (13) does not exist without the integral in Equation (12). The sequence $\{\phi_j(z)\}$ is said to be weakly convergent to $\tilde{\delta}(z)$ [2]. We will use Eq. (13) as a convenient short-hand notation for the actual definition of $\tilde{\delta}(z)$ in Eq. (12).

An alternative derivation of the sifting property of Eq. (2) that does not make use of contour integrals is given in Appendix C. This derivation provides some additional insight into the properties of the generalized delta function.

*2.2 Integral representation*

In the previous section, we represented the generalized delta function by a sequence of Gaussian functions with a complex argument. Here we show that the generalized delta function also has an integral representation that is analogous to that of a conventional Dirac delta function.

Consider the sequence $\{\phi'_j(z)\}$ given by

$$\phi'_j(z) = \frac{1}{2\pi} \int_{-\infty}^{\infty} e^{-izp} e^{-p^2/2\sigma_j'^2} dp, \tag{14}$$

where $\sigma'_j = j$. In analogy with Eq. (12), we define the generalized delta function by

$$\int_{-\infty}^{\infty} f(x)\tilde{\delta}(x - z_0) dx \equiv \lim_{j \to \infty} \int_{-\infty}^{\infty} f(x)\phi'_j(x - z_0) dx. \tag{15}$$

The sifting property of this definition of $\tilde{\delta}(z)$ can be established by simply performing the integral over $p$ in Equation (14), which gives

$$\phi'_j(z) = \frac{\sigma'}{\sqrt{2\pi}} e^{-z^2 \sigma'^2/2}. \tag{16}$$

But $\sigma'_j = 1/\sigma_j$, and a comparison of Equations (4) and (16) shows that $\phi'_j(z) = \phi_j(z)$. Thus the definition of $\tilde{\delta}(z)$ in Equation (15) is identical to that in Equation (12) and all the same properties hold, including the sifting property of Equation (2).

Once again, the symbolic notation

$$\tilde{\delta}(z) = \lim_{\sigma' \to \infty} \frac{1}{2\pi} \int_{-\infty}^{\infty} e^{-izp} e^{-p^2/2\sigma'^2} dp \tag{17}$$

or

$$\tilde{\delta}(z) = \frac{1}{2\pi} \int_{-\infty}^{\infty} e^{-izp} dp \qquad (18)$$

is commonly used [2] for the conventional Dirac delta function, despite the fact that the limit in Equation (17) and the integral in Equation (18) do not exist. We will use Equation (18) as a convenient short-hand notation for the actual definition in Equation Eq. (15). A different proof of Eq. (18) based on Fourier transforms was suggested in Ref. [15].

*2.3 Integrals over the real and imaginary parts of an argument*

Some of the applications of the generalized delta function in quantum optics will involve separate integrals over the real and imaginary parts of a function. In that case, $\tilde{\delta}(z)$ has the property that

$$\int_{-\infty}^{\infty} f(\alpha_r, \alpha_i) \tilde{\delta}(\alpha_r - (x+iy)) d\alpha_r = f(x+iy, \alpha_i),$$
$$\int_{-\infty}^{\infty} f(\alpha_r, \alpha_i) \tilde{\delta}(\alpha_i - (x+iy)) d\alpha_i = f(\alpha_r, x+iy). \qquad (19)$$

Here $\alpha_r$ and $\alpha_i$ are the real and imaginary parts of the argument of a complex analytic function $f(\alpha)$. These properties follow from those of the previous sections provided that the functional form of $f(\alpha_r, \alpha_i)$ can be viewed as an analytic function of $\alpha_r$ and $\alpha_i$ when those parameters are extended into the complex plane. The extension of $\alpha_r$ and $\alpha_i$ into the complex plane has previously been used in the positive-P representation, for example, as is discussed in more detail in the next section. Integrals of this kind will appear in section 4 where we calculate the Glauber-Sudarshan P-function for a cat state, for example.

The results of this section establish the fundamental sifting property of the generalized delta function defined in equations (12) or (15). As mentioned above, reference [16] suggested an entirely different definition for delta functions with a complex argument. As a result, we have introduced the term generalized delta function to eliminate any possible confusion.

## 3. Quasiprobability Distributions

We briefly review some of the properties of quasiprobability distributions in this section. Although these properties are well known, the inclusion of this background information may be beneficial for those readers who are not familiar with quantum optics. As mentioned above, plots of quasiprobability distributions can be very useful in visualizing the nature of nonclassical states [8-9, 11]. Quasiprobability distributions are also an essential tool in many calculations in quantum optics [7,10-11].

One of the most commonly used quasiprobability distributions is the Wigner distribution $W(x,p)$ defined as [4,5]

$$W(x,p) \equiv \frac{1}{\pi} \int_{-\infty}^{\infty} \langle x+q | \hat{\rho} | x-q \rangle e^{-2ipq} dq. \qquad (20)$$

Here $x$ and $p$ are two conjugate quantum variables, such as the two quadratures of a single-mode electromagnetic field. The Wigner distribution has the property that it can have negative values, unlike a true probability distribution.

Negative values of the Wigner distribution often arise from the superposition of two quantum states [11], and the negative regions of the Wigner distribution indicate nonclassical behavior.

An example of a Wigner distribution with negative values is shown in figure 3, which corresponds to a plot of $W(x,p)$ for a number state containing two photons. Although $W(x,p)$ has negative values and thus cannot correspond to a true probability distribution, it has the property that the marginal probability distribution $P_m(x)$ for the variable x is given by [4]

$$P_m(x) = \int_{-\infty}^{\infty} W(x,p)dp. \tag{21}$$

Equation (21) would hold in classical probability theory if $W(x,p)$ were the joint probability distribution for $x$ and $p$. The nonclassical nature of this state requires, however, that $W(x,p)$ take on negative values. A similar relation holds for the marginal probability $P_m(p)$ for the variable $p$.

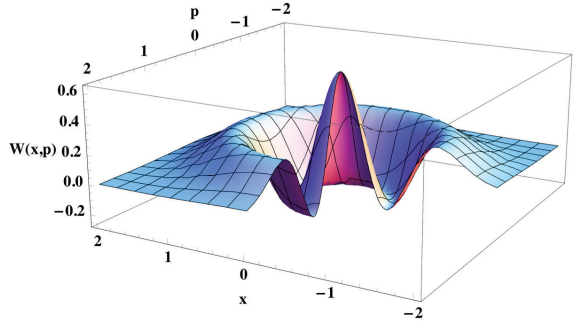

**Figure 3.** The Wigner distribution $W(x,p)$ for the number state $|\psi\rangle = |2\rangle$, plotted as a function of the quadratures $x$ and $p$. A portion of the plot has been removed in order to show the region where the Wigner distribution becomes negative, which illustrates the nonclassical nature of this state.

Two other commonly used quasiprobability distributions are the Glauber-Sudarshan P-function $P(\alpha)$ [4-6,12] and the Husimi-Kano Q-function $Q(\alpha)$ [4-5]. The P-function is implicitly defined in Eq. (1), while the Q-function is defined by

$$Q(\alpha) = \frac{1}{\pi} \langle \alpha | \hat{\rho} | \alpha \rangle. \tag{22}$$

Here $\alpha$ is a complex number corresponding to the amplitude and phase of a coherent state. $Q(\alpha)$ has the advantage that it can be readily calculated for a given density operator directly from equation (22). The Q-function can, in turn, be used to calculate the P-function, which has the advantage that the density operator can be readily calculated from it using equation (1). Although the P-function tends to be highly singular for quantum mechanical systems, $P(\alpha)$ for a Schrödinger cat state can be expressed in a simple form using generalized delta functions as we will show in section 4. By convention, $\alpha$ is related to the real quadrature parameters $x$ and $p$ by

$$\alpha = \frac{1}{\sqrt{2}}(x+ip). \tag{23}$$

More generally, each of these quasiprobability distributions can be calculated from one another [4-5] as illustrated in figure 4. Equations (1) and (22) can be combined and inverted to give [4]

$$P(\alpha) = \frac{1}{(2\pi)^2} \int_{-\infty}^{\infty} \int_{-\infty}^{\infty} e^{|\xi|^2/4} \tilde{Q}(\xi) e^{i(\alpha_r \xi_r + \alpha_i \xi_i)} d\xi_r d\xi_i, \tag{24}$$

where $\tilde{Q}(\xi)$ is the Fourier transform of the Q-function and $\xi = \xi_r + i\xi_i$ is the conjugate variable of $\alpha$. Equation (24) acts as a transformation from the Q-function to the P-function, which corresponds to one of the arrows in figure 3.

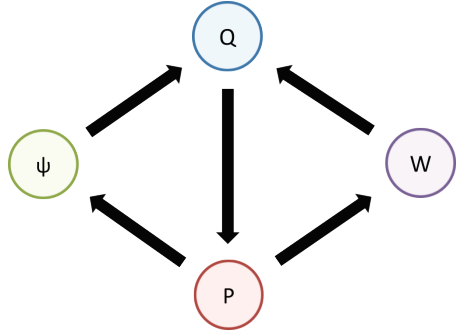

**Figure 4.** Possible transformations from one quasiprobability distribution to another. Here ψ represents the state of the system or the corresponding density matrix for a mixed state, Q represents the Q-function, P represents the P-function and W represents the Wigner distribution. The arrows represent possible transformations from one quasiprobability distribution to another. Other equivalent transformations also exist but are not included here.

The Wigner distribution can be found from the P-function by convolving it with a Gaussian [5]

$$W(\alpha) = \frac{2}{\pi} \int_{-\infty}^{\infty} \int_{-\infty}^{\infty} P(\beta) e^{-2|\alpha-\beta|^2} d\beta_r d\beta_i. \tag{25}$$

The Q-function can then be calculated from the Wigner distribution by convolving it with yet another Gaussian [5]

$$Q(\alpha) = \frac{2}{\pi} \int_{-\infty}^{\infty} \int_{-\infty}^{\infty} W(\beta) e^{-2|\alpha-\beta|^2} d\beta_r d\beta_i. \tag{26}$$

Convolving the P-function with a Gaussian smooths the function so as to eliminate any singularities in the Wigner function. The function is further smoothed when the Wigner function is convolved with yet another Gaussian in equation (26) to obtain the Q-function. Convolving with a Gaussian also tends to smear out some of the quantum mechanical features that can be seen in a plot of the Wigner function, making them less visible in a plot of the Q-function. In those cases where the P-function can be plotted, it tends to display more information than can be seen even in the Wigner function.

The singularities that occur in the Glauber-Sudarshan P-function are due to the fact that it corresponds to a diagonal representation of the density matrix. These singularities can be avoided by using the positive P-function $P_+(\alpha, \beta)$ implicitly defined [17, 18] by

$$\hat{\rho} = \int P_+(\alpha,\beta) \frac{|\alpha\rangle\langle\beta^*|}{\langle\beta^*|\alpha\rangle} d^2\alpha \, d^2\beta. \tag{27}$$

Here $d^2\alpha = d\alpha_r d\alpha_i$ and similarly for $d^2\beta$, while the integral is taken over all space in each of these four variables. It can be seen that the positive P-function corresponds to a non-diagonal expansion in coherent states, which can avoid the singularities inherent in a diagonal expansion. As a result, it is straightforward to calculate $P_+(\alpha,\beta)$ for a Schrödinger cat state. Although the positive P-function is very useful for calculating photon number distributions, for example, our main focus in this paper is the use of generalized delta-functions to describe the singular nature of the diagonal Glauber-Sudarshan P-function.

The positive P-function can be interpreted as extending the real and imaginary parts of the variable $\alpha$ in $P(\alpha)$ into the complex plane [17-18]. As a result, the positive P-function approach is somewhat analogous to Equation (19), where the real and imaginary parts were also extended into the complex plane.

Equations (1), (22) and (24)-(26) all correspond to the transformations shown diagrammatically in figure 4. As mentioned in the caption of figure 4, these transformations are not the only ones that exist but they represent the simplest set of transformations between the various quasiprobability distributions.

## 4. Schrödinger Cat States

Having derived the properties of generalized delta functions and reviewed the theory of quasiprobability distributions, we will now use these results to investigate the singular nature of the Glauber-Sudarshan P-function. As an example, we will derive $P(\alpha)$ for a Schrödinger cat state. This will be done by first calculating $Q(\alpha)$ and then transforming that into $P(\alpha)$, which corresponds to two of the arrows in figure 4. As a consistency check on the results, we will also calculate the density matrix corresponding to $P(\alpha)$ using equation (1) and compare it with the original density matrix, which completes a closed loop of transformations in figure 4.

*4.1 Existence of the P-function*

Because of its singular nature, $P(\alpha)$ does not exist as a true function for highly nonclassical states such as a Schrödinger cat state [5,13]. It has been shown, however, that the P-function can always be represented by a distribution [12], or generalized function. Roughly speaking, this means that the P-function inside an integral can be written as the limit of a sequence of functions $P_M(\alpha)$ where $P_M(\alpha) \to P(\alpha)$ as $M \to \infty$. For any physical density matrix, the functions $P_M(\alpha)$ are infinitely differentiable and decrease faster than $1/|\alpha|^n$ for large $|\alpha|$ and any integer $n$ [12]. It can be shown that the relevant "test functions" $f(z)$ satisfy the conditions assumed in section 1 provided that the energy $E$ of the system is assumed to have an upper bound, which is always the case physically.

To the best of our knowledge, there has only been one previous calculation of $P(\alpha)$ for a Schrödinger cat state $|\psi\rangle$, which we define as

$$|\psi\rangle = A(|\alpha_1\rangle + \zeta|\alpha_2\rangle). \tag{28}$$

Here $\alpha_1$, $\alpha_2$, and $\zeta$ are complex numbers while $A$ is a normalization constant. As shown in ref. [19], the corresponding P-function can be written in the form

$$P(\alpha) = A^2 \Big[ \delta^2(\alpha - \alpha_1) + |\zeta|^2 \delta^2(\alpha - \alpha_2)$$
$$+ e^{|\alpha|^2 - (|\alpha_1|^2 + |\alpha_2|^2)/2} \Big( \zeta^* e^{(1/2)(\alpha_1^* - \alpha_2^*)\partial/\partial(\alpha^* - (\alpha_1^* + \alpha_2^*)/2)} e^{(1/2)(\alpha_2 - \alpha_1)\partial/\partial(\alpha - (\alpha_1 + \alpha_2)/2)} \delta^2(\alpha - (\alpha_1 + \alpha_2)/2) \quad (29)$$
$$+ \zeta e^{(1/2)(\alpha_2^* - \alpha_1^*)\partial/\partial(\alpha^* - (\alpha_1^* + \alpha_2^*)/2)} e^{(1/2)(\alpha_1 - \alpha_2)\partial/\partial(\alpha - (\alpha_1 + \alpha_2)/2)} \delta^2(\alpha - (\alpha_1 + \alpha_2)/2) \Big) \Big].$$

Equation (29) involves exponentials of differential operators and corresponds to an infinite expansion in derivatives of the Dirac delta function. Aside from its complexity, the convergence of this expansion is not apparent and it is of limited usefulness. For example, it would be difficult to use equation (29) to calculate the original state of a Schrödinger cat as a consistency check.

The definition of $P(\alpha)$ in equation (1) corresponds to a diagonal representation of the density operator in a basis of coherent states, whereas the density matrix of a Schrödinger cat state has large off-diagonal terms. A representation of this kind is possible because of the over-completeness of the coherent states [12]. One advantage of our approach is that the use of generalized delta functions explicitly shows how the diagonal $P(\alpha)$ can represent off-diagonal density operators. In particular, the reconstruction of the density operator will involve integrals of the form

$$\int_{-\infty}^{\infty} \int_{-\infty}^{\infty} f(\alpha_r, \alpha_i) \tilde{\delta}(\alpha_i - z_2) \tilde{\delta}(\alpha_r - z_1) d\alpha_r d\alpha_i, \quad (30)$$

which can be evaluated using equation (19).

*4.2 Q-function and its Fourier transform*

The density operator corresponding to the Schrödinger cat state of equation (28), a pure state, is given by

$$\hat{\rho} = A^2 \left( |\alpha_1\rangle\langle\alpha_1| + |\zeta|^2 |\alpha_2\rangle\langle\alpha_2| + \zeta |\alpha_2\rangle\langle\alpha_1| + \zeta^* |\alpha_1\rangle\langle\alpha_2| \right). \quad (31)$$

Each of the four terms in the density operator of equation (31) can be written in general as

$$\rho_{\beta\gamma} = \kappa |\gamma\rangle\langle\beta|. \quad (32)$$

Here $\kappa$ is an appropriate complex constant and $\beta$ and $\gamma$ are equal to either $\alpha_1$ or $\alpha_2$ as appropriate. Since the transformations of equations (1), (22) and (24)-(26) are linear, we can calculate each term of the various quasiprobability distributions for the $\rho_{\beta\gamma}$ separately and combine the results at the end.

Inserting equation (31) into the definition of the Q-function in equation (22) gives

$$Q_{\beta\gamma}(\alpha) = \frac{\kappa}{\pi} \langle \alpha | \gamma \rangle \langle \beta | \alpha \rangle. \quad (33)$$

Using the fact that

$$\langle \alpha | \beta \rangle = e^{-\frac{1}{2}(|\alpha|^2 + |\beta|^2 - 2\alpha^*\beta)}, \quad (34)$$

we can write the general term in the Q-function as

$$Q_{\beta\gamma}(\alpha) = \frac{\kappa}{\pi} e^{-\frac{1}{2}(|\beta|^2+|\gamma|^2)} e^{-|\alpha|^2+\beta^*\alpha+\alpha^*\gamma}. \tag{35}$$

This can be rewritten as

$$Q_{\beta\gamma}(\alpha) = \frac{\kappa}{\pi} e^{-\frac{1}{2}(|\beta|^2+|\gamma|^2-2\beta^*\gamma)} e^{-(|\alpha|^2+\beta^*\gamma-(\beta^*\alpha+\alpha^*\gamma))}. \tag{36}$$

Equation (34) can be used once again to give

$$Q_{\beta\gamma}(\alpha) = \frac{\kappa}{\pi} \langle \beta | \gamma \rangle e^{-(|\alpha|^2+\beta^*\gamma-(\beta^*\alpha+\alpha^*\gamma))}. \tag{37}$$

This gives the full Q-function as

$$Q(\alpha) = \frac{A^2}{\pi} \Big[ e^{-|\alpha-\alpha_1|^2} + |\zeta|^2 e^{-|\alpha-\alpha_2|^2} + \zeta \langle \alpha_1 | \alpha_2 \rangle e^{-(|\alpha|^2+\alpha_1^*\alpha_2-(\alpha_1^*\alpha+\alpha^*\alpha_2))} \\ + \zeta^* \langle \alpha_2 | \alpha_1 \rangle e^{-(|\alpha|^2+\alpha_2^*\alpha_1-(\alpha_2^*\alpha+\alpha^*\alpha_1))} \Big]. \tag{38}$$

Equation (38) corresponds to the Q in figure 4 for the Schrödinger cat state of equation (28).

A calculation of the corresponding P-function will require the Fourier transform of the Q-function, which is given by [4]

$$\tilde{Q}(\xi) \equiv \int_{-\infty}^{\infty} \int_{-\infty}^{\infty} Q(\alpha) e^{-i(\alpha_r\xi_r+\alpha_i\xi_i)} d\alpha_r d\alpha_i. \tag{39}$$

Inserting the value of $Q(\alpha)$ from equation (37) gives

$$\tilde{Q}_{\beta\gamma}(\xi) = \frac{\kappa \langle \beta | \gamma \rangle}{\pi} e^{-\beta^*\gamma} \int_{-\infty}^{\infty} \int_{-\infty}^{\infty} e^{-|\alpha|^2+\beta^*\alpha+\alpha^*\gamma} e^{-i(\alpha_r\xi_r+\alpha_i\xi_i)} d\alpha_r d\alpha_i. \tag{40}$$

The integral in equation (40) can be evaluated to give

$$\int_{-\infty}^{\infty} \int_{-\infty}^{\infty} e^{-|\alpha|^2+\beta^*\alpha+\alpha^*\gamma} e^{-i(\alpha_r\xi_r+\alpha_i\xi_i)} d\alpha_r d\alpha_i = \pi e^{\beta^*\gamma} e^{-|\xi|^2/4} e^{-i(\beta^*+\gamma)\xi_r/2} e^{(\beta^*-\gamma)\xi_i/2}. \tag{41}$$

Inserting equation (41) into (40) gives the expression for $\tilde{Q}_{\beta\gamma}(\xi)$ as

$$\tilde{Q}_{\beta\gamma}(\xi) = \kappa e^{-|\xi|^2/4} \langle \beta | \gamma \rangle e^{-i(\beta^*+\gamma)\xi_r/2} e^{(\beta^*-\gamma)\xi_i/2}. \tag{42}$$

*4.3 Calculation of the P-function*

The P-function for a Schrödinger cat state can now be calculated by inserting the Fourier transform of the Q-function from equation (42) into equation (24), which gives

$$P_{\beta\gamma}(\alpha) = \kappa \langle \beta | \gamma \rangle \left( \frac{1}{2\pi} \int_{-\infty}^{\infty} e^{-i(\beta^*+\gamma)\xi_r/2} e^{i\alpha_r \xi_r} d\xi_r \right) \left( \frac{1}{2\pi} \int_{-\infty}^{\infty} e^{(\beta^*-\gamma)\xi_i/2} e^{i\alpha_i \xi_i} d\xi_i \right). \tag{43}$$

It can be seen that equation (43) involves the inverse Fourier transform of complex exponentials. Using the integral representation of a generalized delta function in equation (18) while making an appropriate change of variables allows equation (43) to be rewritten as

$$P_{\beta\gamma}(\alpha) = \kappa \langle \beta | \gamma \rangle \tilde{\delta}\left( \alpha_r - \frac{\beta^*+\gamma}{2} \right) \tilde{\delta}\left( \alpha_i - i\frac{\beta^*-\gamma}{2} \right). \tag{44}$$

Note that $\gamma = \beta$ in the first two terms in equation (31), in which case

$$P_{\beta\beta}(\alpha) = \kappa \delta(\alpha_r - \text{Re}\{\beta\}) \delta(\alpha_i - \text{Im}\{\beta\}), \tag{45}$$

where $\delta$ is again the usual Dirac delta function. Using the notation [4] following equation (3) gives

$$\delta(\alpha_r - \text{Re}\{\beta\})\delta(\alpha_i - \text{Im}\{\beta\}) = \delta^2(\alpha - \beta). \tag{46}$$

Combining all four terms allows the full P-function for the Schrödinger cat of equation (28) to be written as

$$\begin{aligned} P(\alpha) = \\ A^2 \Bigg[ \delta^2(\alpha - \alpha_1) + |\zeta|^2 \delta^2(\alpha - \alpha_2) + \zeta \langle \alpha_1 | \alpha_2 \rangle \tilde{\delta}\left( \alpha_r - \frac{\alpha_1^* + \alpha_2}{2} \right) \tilde{\delta}\left( \alpha_i - i\frac{\alpha_1^* - \alpha_2}{2} \right) \\ + \zeta^* \langle \alpha_2 | \alpha_1 \rangle \tilde{\delta}\left( \alpha_r - \frac{\alpha_2^* + \alpha_1}{2} \right) \tilde{\delta}\left( \alpha_i - i\frac{\alpha_2^* - \alpha_1}{2} \right) \Bigg], \end{aligned} \tag{47}$$

which is simpler and more useful than equation (29). Equation (47) is one of the main results of this paper.

*4.4 Consistency of the results*

As a consistency check on these results, we will now use the P-function of equation (47) to calculate the corresponding density operator and verify that it is equal to the original density operator. In particular, we would like to understand in more detail how the diagonal $|\alpha\rangle\langle\alpha|$ terms in $P(\alpha)$ in equation (1) can represent off-diagonal density operators.

If the generalized delta functions in equation (47) had the same properties as conventional Dirac delta functions, then their effect inside the integral would be to make the replacement $\alpha \to \alpha'$ where $\alpha'$ is another complex number. In that case, the density operator of equation (1) would have to remain diagonal because

$$|\alpha\rangle \to |\alpha'\rangle, \qquad \langle\alpha| \to \langle\alpha'|. \tag{48}$$

But the effects of the generalized delta function are not at all equivalent to equation (48), as can be seen using the definition of a coherent state in the form

$$|\alpha\rangle = e^{-|\alpha|^2/2} \sum_{n=0}^{\infty} \frac{(\alpha_r + i\alpha_i)^n}{\sqrt{n!}} |n\rangle. \tag{49}$$

Here $|n\rangle$ is a number state containing $n$ photons. The adjoint of this equation gives

$$\langle\alpha| = e^{-|\alpha|^2/2} \sum_{n=0}^{\infty} \frac{(\alpha_r - i\alpha_i)^n}{\sqrt{n!}} \langle n|. \tag{50}$$

For now, we will consider only the third term in equation (47), which has the effect of making the replacement

$$\begin{aligned}
\alpha_r + i\alpha_i &\to (\alpha_1^* + \alpha_2)/2 + i[i(\alpha_1^* - \alpha_2)/2] = \alpha_2 \\
\alpha_r - i\alpha_i &\to (\alpha_1^* + \alpha_2)/2 - i[i(\alpha_1^* - \alpha_2)/2] = \alpha_1^*.
\end{aligned} \tag{51}$$

Aside from a normalizing constant, we see from equations (49)-(51) that

$$|\alpha\rangle \to |\alpha_2\rangle, \qquad \langle\alpha| \to \langle\alpha_1|, \tag{52}$$

in contrast with equation (48). The point is that the generalized delta function has different effects on $\alpha$ and $\alpha^*$, as can be seen from equation (52). The net effect of this is to convert diagonal terms in the density operator of equation (1) into off-diagonal terms.

In more detail, we can rewrite the diagonal terms in equation (1) using

$$|\alpha\rangle\langle\alpha| = e^{-(\alpha_r^2 + \alpha_i^2)} \sum_{j=0}^{\infty} \sum_{k=0}^{\infty} \frac{(\alpha_r + i\alpha_i)^j (\alpha_r - i\alpha_i)^k}{\sqrt{j!k!}} |j\rangle\langle k|. \tag{53}$$

Inserting equations (44) and (53) into the definition of the P-function gives each of the terms $\hat{\rho}_{\beta\gamma}$ in the form

$$\hat{\rho}_{\beta\gamma} = \kappa \langle \beta | \gamma \rangle e^{-\left((\beta^* + \gamma)^2/4 + i^2(\beta^* - \gamma)^2/4\right)} \sum_{j=0}^{\infty} \sum_{k=0}^{\infty} \left(\frac{\beta^* + \gamma + i^2(\beta^* - \gamma)}{2}\right)^j \left(\frac{\beta^* + \gamma + i(-i)(\beta^* - \gamma)}{2}\right)^k \frac{|j\rangle\langle k|}{\sqrt{j!k!}}. \tag{54}$$

Equation (54) can be simplified to give

$$\hat{\rho}_{\beta\gamma} = \kappa \langle \beta | \gamma \rangle e^{-\beta^*\gamma} \sum_{j=0}^{\infty} \sum_{k=0}^{\infty} \frac{\gamma^j \beta^{*k}}{\sqrt{j!k!}} |j\rangle\langle k|. \tag{55}$$

Using equation (34) in equation (55) gives

$$\hat{\rho}_{\beta\gamma} = \kappa e^{-(|\beta|^2+|\gamma|^2)/2} \sum_{j=0}^{\infty} \sum_{k=0}^{\infty} \frac{\gamma^j \beta^{*k}}{\sqrt{j!k!}} |j\rangle\langle k|. \tag{56}$$

From the definition in equations (49) and (50)

$$|\gamma\rangle\langle\beta| = e^{-(|\beta|^2+|\gamma|^2)/2} \sum_{j=0}^{\infty} \sum_{k=0}^{\infty} \frac{\gamma^j \beta^{*k}}{\sqrt{j!k!}} |j\rangle\langle k|. \tag{57}$$

Comparing equations (56) and (57) shows that

$$\hat{\rho}_{\beta\gamma} = \kappa |\gamma\rangle\langle\beta|. \tag{58}$$

Using the appropriate variables and constants in equation (58) gives a density operator that is exactly the same as the original density operator in equation (31).

This result shows that the P-function does give back the correct off-diagonal density matrix when following the cycle illustrated by the arrows in figure 4, as would be expected. Nevertheless, this is a striking result as can be seen by considering a cat state defined by

$$|\psi\rangle = A(|\alpha_0\rangle + e^{i\phi}|-\alpha_0\rangle), \tag{59}$$

where $\alpha_0$ is now a real number. What equation (52) shows in this example is that, when transforming from the P-function back to the original state, the generalized delta functions replace $\langle\alpha|$ with $\langle-\alpha_0|$ while $|\alpha\rangle$ is replaced with $|\alpha_0\rangle$ itself, even though $\alpha_0$ is a real number.

It is worth noting that the P-function for a Schrödinger cat state as given in Eq. (44) corresponds to a sequence of functions of the form

$$P_{\beta,\gamma,M}(\alpha) = \kappa\langle\beta|\gamma\rangle \frac{M^2}{2\pi} \exp\left(\left(\alpha_r - \frac{\beta^*+\gamma}{2}\right)^2 \frac{M^2}{2}\right) \exp\left(\left(\alpha_i - i\frac{\beta^*-\gamma}{2}\right)^2 \frac{M^2}{2}\right). \tag{60}$$

Here we have taken $M = 1/\sigma$ and the distribution representing $P(\alpha)$ corresponds to the limit $M \to \infty$. Each of these functions is infinitely differentiable and they drop off exponentially at infinity since $\alpha_r$ and $\alpha_i$ are both real. Thus our expression for $P(\alpha)$ is consistent with the general properties derived in reference [12].

The positive P-function $P_+(\alpha,\beta)$ can also be readily calculated for a Schrödinger cat state, which can give a non-singular result because $P_+(\alpha,\beta)$ is non-diagonal. The positive P-function can be viewed as extending the real and imaginary parts of $\alpha$ (both of which are real numbers) into the complex plane [17]. That is somewhat similar to our approach in which the argument of the Dirac delta function, which is normally a real number as well, is extended into the complex plane. The two approaches are quite different but they do have this feature in common.

Although we have only considered the use of the generalized delta function to calculate the P-function of a Schrödinger cat state here, similar techniques can be used to calculate the P-function for more general states. The relevance of generalized delta functions to the theory of quantum noise in amplifiers will be discussed in the next section.

## 5. Linear Amplification

In this section we consider a situation in which a Schrödinger cat state is passed through a linear phase-insensitive amplifier with gain $g$. This produces a mixed state with a non-singular $P(\alpha)$ that can be expressed as a sum of Gaussians with a finite width $\sigma$. As the gain approaches unity, $\sigma \to 0$ and these Gaussians become generalized delta functions as described in the previous section. These results show that generalized delta functions can appear naturally in physical systems in the limit of small decoherence.

It has previously been shown by Caves et al. [7] that the effect of a linear phase-insensitive amplifier on any optical signal is to simply scale the parameter $\alpha$ in the Q-function by

$$Q_{out}(\alpha) = \frac{1}{g^2} Q_{in}\left(\frac{\alpha}{g}\right). \tag{61}$$

Here $Q_{in}(\alpha)$ and $Q_{out}(\alpha)$ are the Q-functions before and after the amplification process. This is a surprisingly simple result that allows the effects of quantum noise to be analyzed in a straightforward way.

Applying this transformation to the general Q-function term of a Schrödinger cat in equation (37) gives

$$Q_{\beta\gamma}(\alpha) = \frac{\kappa\langle\beta|\gamma\rangle}{\pi g^2} e^{-\frac{1}{g^2}(|\alpha|^2 + g^2\beta^*\gamma - g(\beta^*\alpha + \alpha^*\gamma))}. \tag{62}$$

Inserting equation (62) into the Fourier transform in equation (39) gives

$$\tilde{Q}_{\beta\gamma}(\xi) = \frac{\kappa\langle\beta|\gamma\rangle}{\pi g^2} e^{-\beta^*\gamma} \int_{-\infty}^{\infty}\int_{-\infty}^{\infty} e^{-\frac{|\alpha|^2}{g^2} + \frac{1}{g}(\beta^*\alpha + \alpha^*\gamma)} e^{-i(\alpha_r\xi_r + \alpha_i\xi_i)} d\alpha_r d\alpha_i. \tag{63}$$

Evaluating the integral in equation (63) gives

$$\int_{-\infty}^{\infty}\int_{-\infty}^{\infty} e^{-\frac{|\alpha|^2}{g^2} + \frac{1}{g}(\beta^*\alpha + \alpha^*\gamma)} e^{-i(\alpha_r\xi_r + \alpha_i\xi_i)} d\alpha_r d\alpha_i = \pi g^2 e^{\beta^*\gamma} e^{-g^2\frac{|\xi|^2}{4}} e^{-\frac{ig}{2}(\beta^*\xi + \xi^*\gamma)}. \tag{64}$$

Combining equations (63) and (64) gives the Fourier Transform of the Q-function as

$$\tilde{Q}_{\beta\gamma}(\xi) = \kappa\langle\beta|\gamma\rangle e^{-g^2\frac{|\xi|^2}{4}} e^{-\frac{ig}{2}(\beta^*\xi + \xi^*\gamma)}. \tag{65}$$

Equation (65) can now be inserted into equation (24) to obtain the P-function in the form

$$P_{\beta\gamma}(\alpha) = \frac{\kappa\langle\beta|\gamma\rangle}{(2\pi)^2} \int_{-\infty}^{\infty}\int_{-\infty}^{\infty} e^{-(g^2-1)\frac{|\xi|^2}{4}} e^{-\frac{ig}{2}(\beta^*\xi + \xi^*\gamma)} e^{i(\alpha_r\xi_r + \alpha_i\xi_i)} d\xi_r d\xi_i. \tag{66}$$

The integral in equation (66) can be evaluated to give

$$\int_{-\infty}^{\infty}\int_{-\infty}^{\infty} e^{-(g^2-1)\frac{|\xi|^2}{4}} e^{-\frac{ig}{2}(\beta^*\xi + \xi^*\gamma)} e^{i(\alpha_r\xi_r + \alpha_i\xi_i)} d\xi_r d\xi_i = \frac{4\pi}{g^2-1} e^{-\frac{1}{g^2-1}(|\alpha|^2 + g^2\beta^*\gamma - g(\beta^*\alpha + \alpha^*\gamma))}. \tag{67}$$

Inserting equation (67) into equation (66) gives the general term in the P-function as

$$P_{\beta\gamma}(\alpha) = \frac{\kappa\langle\beta|\gamma\rangle}{\pi(g^2-1)} e^{-\frac{1}{g^2-1}(|\alpha|^2 + g^2\beta^*\gamma - g(\beta^*\alpha + \alpha^*\gamma))}. \tag{68}$$

Combining all four terms $P_{\beta\gamma}(\alpha)$ gives the full P-function for the amplified Schrödinger cat state as

$$\begin{aligned}P_{out}(\alpha) = \\ \frac{A^2}{\pi(g^2-1)}\Big[ & e^{-|\alpha - g\alpha_1|^2/(g^2-1)} + |\zeta|^2 e^{-|\alpha - g\alpha_2|^2/(g^2-1)} + \zeta\langle\alpha_1|\alpha_2\rangle e^{-(|\alpha|^2 + g^2\alpha_1^*\alpha_2 - g(\alpha_1^*\alpha + \alpha^*\alpha_2))/(g^2-1)} \\ & + \zeta^*\langle\alpha_2|\alpha_1\rangle e^{-(|\alpha|^2 + g^2\alpha_2^*\alpha_1 - g(\alpha_2^*\alpha + \alpha^*\alpha_1))/(g^2-1)} \Big].\end{aligned} \tag{69}$$

We now define the parameter $\sigma$ by

$$\sigma = \sqrt{\frac{g^2-1}{2}}. \tag{70}$$

Inserting equation (70) into equation (68) allows the four terms in $P(\alpha)$ to be rewritten in the form

$$P_{\beta\gamma}(\alpha) = \kappa\langle\beta|\gamma\rangle \left(\frac{1}{\sqrt{2\pi}\sigma} e^{-(\alpha_r - g(\beta^* + \gamma)/2)^2/2\sigma^2}\right) \left(\frac{1}{\sqrt{2\pi}\sigma} e^{-(\alpha_i - ig(\beta^* - \gamma)/2)^2/2\sigma^2}\right). \tag{71}$$

It can be seen from Equation (13) that the two exponential terms in equation (71) converge weakly [2] to generalized delta functions in the limit as $\sigma \to 0$, which corresponds to taking the limit $g \to 1$. In that limit, equation (69) reduces to the P-function for a Schrödinger cat without amplification as given in equation (47).

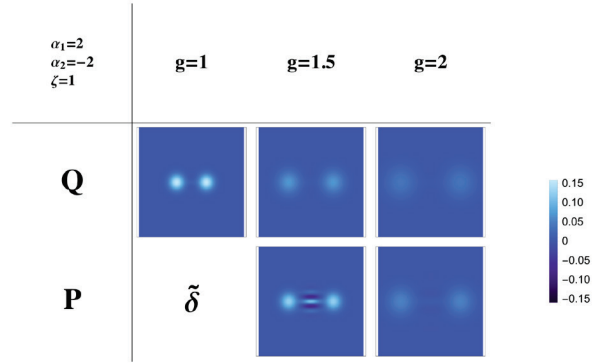

**Figure 5.** Plots of the Q and P functions given by equations (62) and (69) respectively for different values of the gain g. These correspond to a Schrödinger cat state from equation (28) with the parameters shown in the upper-left corner. Note that when g approaches unity the P-function becomes the generalized delta function and cannot be plotted. It can also be seen that the P-function and Q-function become approximately the same when the gain is relatively large.

Plots of $Q(\alpha)$ and $P(\alpha)$ for the amplified Schrödinger cat state are shown in figure 5 for several values of the gain. The P-function cannot be plotted for the case of $g=1$, since it reduces to a combination of generalized delta functions in that limit. The functional form of a generalized delta function in the complex plane is graphically illustrated, however, in Appendix A.

These results show that generalized delta functions can arise naturally in the description of quantum-mechanical systems in the limit of small decoherence, which suggests that they may occur in a number of other applications as well.

## 6. Summary and Conclusions

We have generalized the Dirac delta function to allow the use of complex arguments. The generalized delta function $\tilde{\delta}(z)$ can be defined as the limit of a sequence of Gaussian distributions that is equivalent to one of the usual representations of a Dirac delta function but with a complex argument $z$. An equivalent integral representation was given in Equations (14-18). Although the sifting property and representations of the generalized delta function are the same as that of an ordinary delta function, there are several important differences when using a complex argument. It was shown that $\int f(x)\tilde{\delta}(x-z_0)dx$ along the real axis gives the value $f(z_0)$ provided that the test function $f$ is analytic throughout the complex plane, which is more restrictive than the requirement for an ordinary Dirac delta function. As illustrated in Fig. 1, this property of generalized delta functions is more similar to a contour integral around a pole than it is to a conventional Dirac delta function. In addition, $\tilde{\delta}(z)$ is nonzero and singular over an extended region of the real axis, unlike a conventional Dirac delta function. Delta functions with complex arguments have been briefly discussed previously [14-16] but with several different definitions and no rigorous proof of their properties.

Part of our motivation for considering generalized delta functions is their ability to describe the singular nature of the Glauber-Sudarshan P-function $P(\alpha)$. Here we calculated $P(\alpha)$ for a Schrödinger cat state in a simple form that involves generalized delta functions. As a consistency check on this result, we used $P(\alpha)$ to calculate the corresponding density operator, which agreed with that of the original Schrödinger cat state. This approach clearly shows how the diagonal P-function can represent density operators with off-diagonal terms.

We also showed that the Gaussian representation of the generalized delta function arises naturally when a Schrödinger cat state is passed through a linear phase-insensitive amplifier. The P-function in that case corresponds to a set of Gaussians with a finite width $\sigma$ for gain $g > 1$, while $\sigma \to 0$ in the limit of $g \to 1$. Thus generalized delta functions may be expected to play a role in a variety of physical systems in the appropriate limit. For example, it has previously been noted that delta functions with complex arguments can be used to simplify the analysis of asymptotic integrals encountered in classical electromagnetism [14]. We are currently using generalized delta functions to calculate the P-function of other quantum superposition states and we have found them to be invaluable in analyzing the effects of linear amplifiers on entangled states, for example. Our approach was also recently used to calculate analytical solutions to the Wigner-Boltzmann transport equation [20]. As a result, we expect that generalized delta functions will be of use in a variety of future applications.

## Acknowledgments

This work was supported in part by the National Science Foundation under grant No. 1402708 and by a GAANN fellowship under Department of Education grant #P200A150003.

## Appendix A. Functional Form

The generalized delta function can be highly singular over an extended region of the real axis. Here we consider the functional form of the generalized delta function and provide a graphic illustration of the nature of its singularity.

We begin by considering the Gaussian series representation of the generalized delta function given by equation (13) in the text:

$$\tilde{\delta}(z) \equiv \lim_{\sigma \to 0} \frac{1}{\sigma\sqrt{2\pi}} e^{-z^2/2\sigma^2}. \tag{A1}$$

Let $z = z_r + i z_i$ which gives

$$\tilde{\delta}(z) = \lim_{\sigma \to 0} \frac{1}{\sigma\sqrt{2\pi}} e^{-(z_r^2 - z_i^2)/2\sigma^2} e^{iz_r z_i/\sigma^2}. \tag{A2}$$

If we ignore the complex exponential in equation (A2) for the moment, it is apparent that the limit goes to zero whenever $z_r^2 > z_i^2$ while it diverges when $z_r^2 \leq z_i^2$. Returning to the complex exponential, it can be seen that the imaginary part of $\tilde{\delta}(z)$ is zero when either $z_r$ or $z_i$ is equal to zero, while it is undefined otherwise.

We know that the value of the complex exponential in equation (A2) must lie somewhere on the complex unit circle. The entire expression must go to zero when the first factor in equation (A2) goes to zero. However, the location of the second factor on the complex unit circle becomes undefined when the first exponential term goes to $\infty$. This situation is known as complex infinity, which we denote by $\tilde{\infty}$. Combining these arguments, the functional form of the generalized delta function can be summarized as

$$\tilde{\delta}(z) = \begin{cases} \infty & \text{if } \operatorname{Re}\{z\} = 0 \\ 0 & \text{if } \operatorname{Re}\{z\} \neq 0 \text{ and } \operatorname{Re}\{z\}^2 > \operatorname{Im}\{z\}^2 \\ \tilde{\infty} & \text{if } \operatorname{Re}\{z\} \neq 0 \text{ and } \operatorname{Re}\{z\}^2 \leq \operatorname{Im}\{z\}^2 \end{cases}. \tag{A3}$$

These results are illustrated schematically in figure A1.

By comparison, the usual Dirac delta function can be described by

$$\delta(x) = \begin{cases} \infty & \text{if } x = 0 \\ 0 & \text{if } x \neq 0. \end{cases} \tag{A4}$$

It can be seen that equation (A3) reduces to equation (A4) when the imaginary part of $z$ is equal to zero.

Equation (A3) and figure A1 represent the form of the generalized delta function itself. When it is included in an integral with a complex argument, the center point of figure A1 is shifted to a point in the complex plane. As a result, the integrand is singular over an extended region of the real axis. As discussed in the text, it is understood that $\tilde{\delta}(z)$ is only defined when it is included in such an integral.

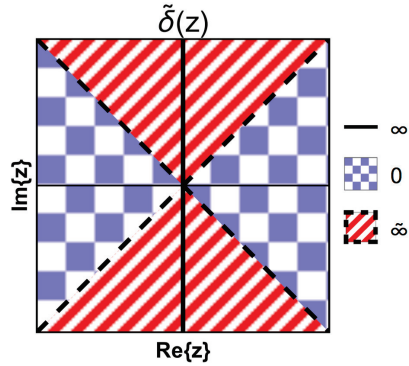

**Figure A1.** Schematic representation of the form of the generalized delta function $\tilde{\delta}(z)$ as summarized in equation (A3). The origin of this plot would be shifted into the complex plane when the generalized delta function is included in an integral with a complex argument.

## Appendix B: Alternative Definitions

As mentioned in the main text, a delta function with a complex argument has been defined in two different ways [14-16]. Our definition of the generalized delta function in equation (2) was motivated in part by the need to distinguish between these two definitions, as is discussed in more detail in this appendix.

A delta function with a complex argument has previously been defined [16] by

$$\delta[(x+iy)-(a+ib)] = \delta[(x-a)+i(y-b)]$$
$$= \delta(x-a)\delta(y-b). \tag{B1}$$

In our notation [4] this would be written instead as $\delta^2$. From equation (B1) one can see that this is actually a product of two ordinary Dirac delta functions. This delta function has a representation given by

$$\delta^2(z_r + iz_i) \equiv \lim_{\sigma \to 0} \frac{1}{2\pi\sigma^2} e^{-z_r^2/2\sigma^2} e^{-z_i^2/2\sigma^2}, \tag{B2}$$

which is quite different from the definition of $\tilde{\delta}(z)$ in equation (13) in the text.

The generalized delta function $\tilde{\delta}(z)$ is used inside one-dimensional integrals, whereas $\delta^2(z)$ is intended to be included in a double integral over both the real and imaginary part of the argument. It has the property that

$$\int_{-\infty}^{\infty}\int_{-\infty}^{\infty} f(\zeta)\delta^2(\zeta - z)d\zeta_r d\zeta_i = f(z). \tag{B3}$$

The Fourier transform representation of $\delta^2(z)$ is given by

$$\delta^2(z_r + iz_i) = \frac{1}{(2\pi)^2}\int_{-\infty}^{\infty}\int_{-\infty}^{\infty} e^{iz_r\xi_r}e^{iz_i\xi_i}d\xi_r d\xi_i, \tag{B4}$$

which can be contrasted with equation (18).

We have defined the generalized delta function $\tilde{\delta}(z)$ in order to distinguish it from the product of two ordinary Dirac delta functions. Both $\tilde{\delta}(z)$ and $\delta^2(z)$ are useful, however, and both appear in our analysis of Schrödinger cat states in the text.

**Appendix C. Alternate Proof of the Sifting Property**

In the main text, the sifting property of equation (2) was proven using the contour integral of figure 2. In this appendix we give an alternate derivation that provides additional insight into the nature of the generalized delta function.

In order to derive equation (2), we make use of the fact that the test function $f(z)$ has been assumed to be analytic throughout the complex plane, which allows it to be expanded in a Taylor series as

$$f(z) = \sum_{n=0}^{\infty} \frac{f^{(n)}(0)}{n!} z^n. \tag{C1}$$

Here $f^{(n)}(z)$ is the $n^{\text{th}}$ derivative of $f$. Inserting equation (C1) into equation (2) gives

$$F(z) = \sum_{n=0}^{\infty} \frac{f^{(n)}(0)}{n!} \int_{-\infty}^{\infty} x^n \tilde{\delta}(x-z)dx, \tag{C2}$$

where we have let $F(z)$ denote the result of the integral. The sum and the integral can be interchanged because equation (C1) is uniformly convergent [2]. If we define the integral $I_n$ in equation (C2) as

$$I_n = \int_{-\infty}^{\infty} x^n \tilde{\delta}(x-z)dx, \tag{C3}$$

then it can be seen that equation (2) will hold provided that $I_n = z^n$.

From the definition of the generalized delta function in equation (12) of the main text, $I_n$ is given as

$$I_n = \lim_{\sigma \to 0} \frac{1}{\sigma\sqrt{2\pi}} \int_{-\infty}^{\infty} x^n e^{-(x-z)^2/2\sigma^2} dx. \tag{C4}$$

Equation (C4) may be rewritten as

$$I_n = \lim_{\sigma \to 0} \frac{e^{-z^2/2\sigma^2}}{\sigma\sqrt{2\pi}} \int_{-\infty}^{\infty} x^n e^{-x^2/2\sigma^2} e^{xz/\sigma} dx. \tag{C5}$$

It can be shown that

$$\int_{-\infty}^{\infty} x^n e^{-ax^2+bx} dx = \sqrt{\frac{\pi}{a}} \left(-\frac{i}{2\sqrt{a}}\right)^n H_n\left(\frac{ib}{2\sqrt{a}}\right), \tag{C6}$$

for all $a > 0$ and complex $b$, where $H_n$ are the Hermite polynomials. Using equation (C6), the integral in equation (C5) can be evaluated to give

$$I_n = \lim_{\sigma \to 0} \frac{f^{(n)}(0)}{n!} \left(-\frac{i\sigma}{\sqrt{2}}\right)^n H_n\left(\frac{iz}{\sqrt{2}\sigma}\right) \tag{C7}$$

The Hermite polynomials can be defined, explicitly, as

$$H_n(x) = \sum_{m=0}^{\lfloor n/2 \rfloor} \frac{(-1)^m}{m!(n-2m)!} (2x)^{n-2m}, \tag{C8}$$

where $\lfloor x \rfloor$ represents the floor function. We can now consider the limit of equation (C7) as $\sigma \to 0$. From equation (C8), the Hermite polynomials in equation (C7) have terms that involve $\sigma^k$ where k is an integer less than or equal to n. The factor outside the Hermite polynomial in equation (C7) involves $\sigma^n$. As a result, the only terms in equation (C7) that do not go to 0 as $\sigma \to 0$ are the ones with $k = n$. Using this fact reduces equation (C7) to

$$I_n = z^n. \tag{C9}$$

Inserting equations (C9) and (C5) into equation (C2) gives the Taylor series expansion of equation (C1), which shows that $F(z) = f(z)$ as desired.

It can be seen from equations (C5) and (C9) with $n = 0$ that

$$\int_{-\infty}^{\infty} \tilde{\delta}(x-z)dx = 1. \tag{C10}$$

Equation (C10) demonstrates that the generalized delta function is a properly normalized distribution.

**References**

[1] S. G. Georgiev, *Theory of Distributions* (Springer, Berlin, 2015).
[2] E. Butkov, *Mathematical Physics* (Addison-Wesley, Reading, MA, 1968).
[3] M.J. Lighthill, *An Introduction to Fourier Analysis and Generalised Functions* (Cambridge University Press, Cambridge, 1958).
[4] W.P. Schleich, *Quantum Optics in Phase Space* (WILEY-VCH, Berlin, 2001).
[5] K.E. Cahill and R.J. Glauber, Phys. Rev. **177**, 1882-1902 (1969).
[6] E.C.G. Sudarshan Phys. Rev. Lett. **10**, 277-279 (1963).
[7] C.M. Caves, J. Combes, Z. Jiang, and S. Pandey, Phys. Rev. A. **86**, 063802 (2012).
[8] K. Vogel and H. Risken, Phys. Rev. A. **40**, 2847-2849 (1989).
[9] V. Bužek and M.S. Kim, Phys. Rev. A. **48**, 3394-3397 (1993).
[10] J. Eiselt and H. Risken, Phys. Rev. A. **43**, 346-360 (1991).
[11] C. Ferrie, Rep. Prog. Phys. **74**, 116001 (2011).
[12] J.R. Klauder and E.C.G. Sudarshan, *Fundamentals of Quantum Optics* (W.A. Benjamin, Inc., New York, 1968).
[13] B. Sanders, Phys. Rev. A. **45**, 6811-6815 (1991).
[14] I.V. Lindell, Am. J. Phys. **61**, 438-442 (1993).
[15] A.D. Poularkis, *Transforms and Applications Handbook* (CRC Press, Boca Raton, 2000) 2nd ed.
[16] J.W. Harris and H. Stocker, *Handbook of Mathematics and Computational Science* (Springer-Verlag, New York, 1998).
[17] P.D. Drummond and C.W. Gardiner, J. Phys. A. **13**, 2353-2368 (1980).
[18] M. Wolinsky and H.J. Carmichael, Phys. Rev. Lett. **60**, 1836-1839 (1988).
[19] L. Mandel, Physica Scripta. **12**, 34-42 (1986).
[20] A.R. Fernandes Nt. and L.F. Santos, arXiv:1705.0024 (2017).